\documentclass[twoside,twocolumn,showpacs,aps]{revtex4}
\usepackage{graphicx}
\usepackage{amsmath}
\newcommand{\be}{\begin{equation}}
\newcommand{\ee}{\end{equation}}
\newcommand{\ben}{\begin{eqnarray}}
\newcommand{\een}{\end{eqnarray}}
\begin{document}
\title{Presence of asymmetric noise in multi-terminal chaotic cavities}
\author{A. L. R. Barbosa$^1$, J. G. G. S. Ramos$^2$, and D. Bazeia$^2$}
\affiliation{$^1$Unidade Acad\^emica de Ensino a Dist\^{a}cia e Tecnologia, Universidade Federal Rural de Pernambuco, 52171-900 Recife, PE, Brazil\\
$^2$Departamento de F\'\i sica, Universidade Federal dea Para\'\i ba, 58051-900 Jo\~ao Pessoa, PB, Brazil}
\date{\today}
\begin{abstract}
This work deals with chaotic quantum dot connected to two and four leads. We use standard diagrammatic procedure to integrate on the unitary group, to study the main term in the semiclassical expansion of the noise in the three pure Wigner-Dyson ensembles at finite frequency and temperature, in the noninteracting and interacting regimes. We investigate several limits, related to the temperature and the potential difference in the leads, in the presence of a capacitive environment. At the thermal crossover regime, we obtain general expressions described in terms of parameters that can be controlled experimentally. As an interesting result, we show the appearance of asymmetries in the noise, controlled by the topology of the cavity and the number of open channels in the corresponding terminals.
\end{abstract}
\maketitle
\section{Introduction}

The study of coherent charge transport through chaotic mesoscopic quantum dots is of current interest in physics \cite{blanter,datta,mello,Nazarovbook,mont}. Subtle effects appear from the interference originated through the multiple coherent scattering \cite{{LangenButtiker,PedersenButtiker}}, which strongly depend on external parameters such as the magnetic field, the spin-orbit coupling, the topology, the temperature and other phenomenological parameters that can be controlled experimentally \cite{gustavson}. An important observable of the quantum transport which strongly depends on the finite size of the chaotic quantum dot is the shot noise power, the second cumulant of the counting statistics of the electrons at zero temperature. Experimental measurements of the shot noise power and of other cumulants have been performed in \cite{gustavson1,gustavson2}. The shot noise provides direct information about the discreteness of matter \cite{revNazarov,Whitney,Beri}, so it is different from the thermal noise \cite{blanter,leturcq}. In mesoscopic systems, the two source of noises contribute for the quantum transport of electrons through the cavity at finite temperature.

Nontrivial properties of the noise peaks indicate a subtle combination of both spatial and temporal coherence in the electronic propagation through the system. In chaotic quantum dots, in accordance with the results of Ref.~\cite{chen}, asymmetries in the spectral density of the noise in the Kondo regime are obtained in terms of the tension in the electron reservoirs. In this regime, the asymmetry was shown to strongly depend on the AC frequency which irradiate the chaotic quantum dots connected to two terminals. The presence of the AC frequency also induces asymmetric peaks in the derivative of the noise. Another effect that shows radical change in the shot noise power of a chaotic quantum dot was pointed out in \cite{nos,epl}.

In the Landauer-Buttiker framework, a particularly interesting characteristic of the quantum transport in chaotic mesoscopic systems is the dependence with the geometry. A critical example which shows the importance of the geometry of the chaotic mesoscopic systems appeared in Ref.~{\cite{Texier}}, where the authors identified a change in the correction of the weak localization of the conductance in the case of a multi-terminal network made of quasi-one-dimensional diffusive wires. The effect there identified is directly related to the geometry: for a system of $N$ wires connected to the center of the conducting
quasi-one-dimensional region, the Cooperon correction to the quantum interference reads $\delta T=1/3(N/4-1)$. Since $\delta T$ vanishes for $N=4$, a change of signal is induced geometrically, indicating a transition depletion-enhancement for the conductance due to the quantum interference.

Other recent investigations concerning topology of quantum dots connected to several terminals appeared in \cite{Bardar}. A key issue concerned the time-dependent fluctuations in the electronic current injected from reservoirs with non-equilibrium spin accumulation into mesoscopic conductors. There, using  topology the authors could suggest an interesting way to measure spin accumulation in the regime of vanishing frequency.

The above results motivated us to study issues related to the topology of the mesoscopic system, concerning the number of leads and open channels connected to the chaotic quantum dot. More specifically, in the current work we show the appearance of an asymmetry in the main term in the semiclassical expansion of the noise power, as a function of the potential in each reservoir and the number of open channels in the leads, in terms of parameters that can be controlled experimentally.

We study the chaotic quantum dot depicted in Fig. 1, searching for effects due to the finite temperature, the difference of potential among the leads and the influence of the frequency of an AC current on the noise. In the mesoscopic system, each lead is connected to an electronic reservoir with $\mu_i = eV_i$, $i = 1,2, 3,$ and $4$, at temperature $T$. Here we take $V_{12} = V_1 - V_2$ and $V_{34} = V_3 - V_4$. Thermal effects appear through the presence of temperature in the electron reservoirs, and lead us to interesting scenarios, in which the chaotic cavity gives rise to the crossover between the thermal noise, present in electrical circuits at high temperatures, and the shot noise power, induced by the quantum effects that appear in electronic circuits at very low temperatures \cite{blanter,gustavson,pekola,epl}. Also, one uses $C$ to represent the capacitive environment, which is introduced to generate interaction among the charge carriers inside the quantum dot \cite{blanter,PedersenButtiker}.

The investigation starts using diagrammatic approach to obtain general expressions for the noise power in the multi-terminal chaotic cavity. We use the results to develop specific expressions and, at the very end of the study, we comment on several plotted results that show a surprising phenomenology and a rich manner to measure the noise power. For pedagogical reasons, we organize the work as follows: In the next Sec. II, we introduce the current-current correlation function, using the scattering theory for quantum transport and the noise in terms of the scattering matrix, to describe charge transport through the chaotic quantum dot. We use the diagrammatic procedure to calculate averages over the unitary group \cite{BrouwerBeenakker,nosJPA}. The investigation follows in Sec. III, where we consider the noninteracting case, with vanishing frequency. We consider the interacting case in Sec. IV, where we show explicitly that the equations there obtained for the noise satisfy the current conservation law if one substitutes the dwell time, $\tau_D$, by the charge relaxation time. Finally, we end the investigation in  Sec. V, where we deal with the rich phenomenology of the noise in the multi-terminal case, including the surprising effects of the asymmetry which appears controlled by the extra terminals. The main results of the current work are shown in Figs. 2, 3, 4, 5, 6, 7, and 8, and we end the paper in Sec. VI, where we include our comments and conclusions.

\bigskip
\begin{figure}
\begin{center}
\includegraphics[width=6.0cm,height=6.0cm]{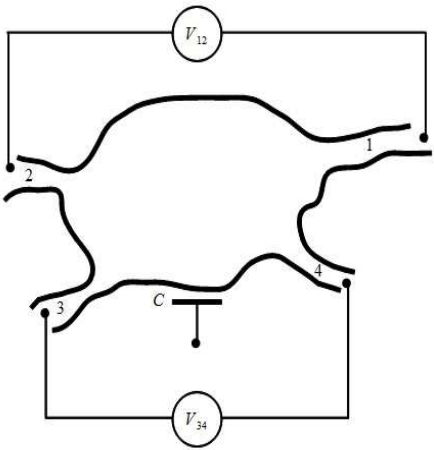}
\end{center}
\caption{Illustration of the chaotic quantum dot connected to four leads, with $C$ standing for the capacitive environment. We use $V_{12}=V_1-V_2$ and $V_{34}=V_3-V_4$ to identify the difference of potentials between the leads 1 and 2, and 3 and 4, respectively.} \label{ponto}
\end{figure}

\section{Scattering theory for quantum transport}\label{sistemaquantico}

We start considering the time-dependent current $\hat{I}_{\gamma}(t)$ at lead $\gamma$, for $\gamma=1,2,...,m$, with $m$ being the number of leads connected to the chaotic quantum dots. Within the framework of the scattering theory for quantum transport, the current-current correlation function can be written in the form \cite{blanter}
\begin{eqnarray}
\langle \delta \hat{I}_\gamma(t) \delta \hat{I}_\alpha(0)\rangle=\int\frac{dw}{2\pi}e^{-iwt}\mathcal{S}_{\gamma \alpha}(w),
\end{eqnarray}
where $ \delta \hat{I}_\alpha(t)\equiv  \hat{I}_\alpha(t)+\langle \hat{I}_\alpha(t)\rangle$ and the noise in the absence of interaction is
\begin{eqnarray}
\mathcal{S}_{\gamma\alpha}(w)&=&\sum_{\nu,\rho}\frac{e^2}{h}\int d\varepsilon\; \text{Tr}\left\lbrace A_{\rho\gamma}(\varepsilon,\varepsilon+\hbar w)\right.\nonumber\\&&\left.\times A_{\nu\alpha}(\varepsilon+\hbar w, \varepsilon)\right\rbrace\nonumber\\&&\times \left\{f_\nu(\varepsilon)\left[1-f_\rho(\varepsilon+\hbar w)\right]\right.\nonumber\\&&\left.+f_\rho(\varepsilon)\left[1-f_\nu(\varepsilon+\hbar w)\right]\right\}.\label{ruido}
\end{eqnarray}
The matrix $A_{\nu\alpha}(\varepsilon, \varepsilon')$ is the current matrix, and it is defined as $ A_{\nu\alpha}(\varepsilon, \varepsilon')=1_\alpha1_\nu-1_\nu S^{\dagger}(\varepsilon)1_\alpha S(\varepsilon')$, where $S(\varepsilon)$ is the scattering matrix, which depends on the energy $\varepsilon$ and describes the charge transport through the circuit. Also, $1_\alpha$ is the projection matrix, which projects onto the lead $\alpha$, and $f_\alpha(\varepsilon)=\left(1+\exp{\left[\beta(\varepsilon-\mu_\alpha)\right]}\right)^{-1}$ represents the Fermi distribution function, related to the thermal reservoir connected to the lead $\alpha$.

The scattering matrix $S(\varepsilon)$ used to describe the mesoscopic system is uniformly distributed over the ortogonal group, if time reversal and spin rotation symmetries are both present in the system, or over the unitary group, if the time reversal symmetry is broken by the action of a strong external magnetic field, or yet over the simpletic group, if the spin rotation symmetry is broken by the action of intense spin-orbit interaction \cite{mello}. To obtain the average of the noise, according to Eq.~\eqref{ruido}, we have to calculate the average of the product of two and four scattering matrices $S(\varepsilon)$. Since we are interested in the main term for the semiclassical expansion of the noise, each ensemble gives the same average value, discarding higher order contributions. As an efficient way to get to such average value, we use the diagrammatic procedure developed in Ref.~\cite{BrouwerBeenakker}.

This diagrammatic procedure requires that we appropriately parametrize the scattering matrix. We follow \cite{nosJPA,bb96}, and we parametrize $S(\varepsilon)$ within the stub model, in a manner which allows including external fields and other relevant phenomenological parameters:
\begin{equation}
S(\varepsilon)=P[1-U Qr(\varepsilon)Q^T]^{-1}UP^T.
\end{equation}
The matrix $U$ is distributed in one of the Wigner-Dyson ensembles, of dimension $M\times M$, where $M=\sum^m_\gamma N_\gamma$. $P$ and $Q$ are projection matrices, $(M+N_s)\times M$ and $M\times N_s$, respectively. Also, $r(\varepsilon)=e^{i\varepsilon \Phi/M}$ is the matrix which describes the stub, coupled to the chaotic quantum dot of dimension $N_s\times N_s$, with $N_s$ standing for the number of open channels in the stub. The matrix $\Phi$ is Hermitian, positive, such that $\phi= \text{Tr}\Phi$. In the reference \cite{BrouwerButtiker}, the authors show that $\phi$ is a parameter related to the average density of modes. Thus, we can associate the stub to specific time scale, to characterize the lifetime of the metastable electronic modes in the chaotic quantum dot, that is, we can write $\phi =\tau_D M/\hbar$, where $\tau_D$ is the dwell time.

In the limit $M \gg 1$, we can expand $S(\varepsilon)$ in power of $U$. Performing the diagrammatic procedure, we can obtain the average of the traces of the two and four scattering matrices and it is possible to verify that only the ladder diagrams (known as difusons) contribute to the main semiclassical term.
The diagrams for the average of the trace of product of two scattering matrices $S(\varepsilon)$ can be found in \cite{BrouwerBeenakker}, while the diagrams for the average of the trace of product of four scattering matrices are found in \cite{nosJPA}. With this, we can obtain for a chaotic quantum dot connected to several leads the following results:

\begin{equation}\label{main2}
\langle\text{Tr}\left [A_{\nu\alpha}(\varepsilon, \varepsilon')\right]\rangle= \delta_{\alpha\nu} N_{\alpha}-\frac{N_\alpha N_\nu}{M\left[1-i\frac{(\varepsilon'-\varepsilon)\tau_D}{\hbar}\right]},
\end{equation}
and
\begin{widetext}
\begin{eqnarray}\label{main4}
\langle\text{Tr}\left[1_\rho S^{\dagger}(\varepsilon)1_\gamma S(\varepsilon')1_\nu S^{\dagger}(\varepsilon'')1_\alpha S(\varepsilon''')\right]\rangle=\frac{N_\gamma N_\rho}{M}\left[\frac{N_\alpha}{\left[1-i\frac{(\varepsilon'-\varepsilon)\tau_D}{\hbar}\right]\left[1-i\frac{(\varepsilon'''-\varepsilon'')\tau_D}{\hbar}\right]}\delta_{\rho \nu}\qquad\qquad\qquad\right.\label{media4sb}\nonumber\\
+\frac{N_\nu}{\left[1-i\frac{(\varepsilon'''-\varepsilon)\tau_D}{\hbar}\right]\left[1-i\frac{(\varepsilon'-\varepsilon'')\tau_D}{\hbar}\right]}\delta_{\gamma \alpha}
\!\!+\!\!\left.\frac{N_\alpha N_\nu}{M}\frac{\left[1-i\frac{(\varepsilon'''+\varepsilon'-\varepsilon''-\varepsilon)\tau_D}{\hbar}\right]}{\left[1\!-\!i\frac{(\varepsilon'-\varepsilon)\tau_D}{\hbar}\right]
\!\!\left[1\!-\!i\frac{(\varepsilon'''-\varepsilon'')\tau_D}{\hbar}\right]\!\!\left[1\!-\!i\frac{(\varepsilon'''-\varepsilon)\tau_D}{\hbar}\right]\!\!\left[1\!-\!i\frac{(\varepsilon'-\varepsilon'')\tau_D}{\hbar}\right]}\right],
\end{eqnarray}
\end{widetext}
where $\alpha,\gamma,\nu,\rho=1,2,...m$. Expressions like (\ref{main2}) and (\ref{main4}) were obtained before in Refs.~[\onlinecite{BrouwerButtiker}] and [\onlinecite{pekola}], respectively, for chaotic quantum dots connected to two leads.

\begin{figure}[ht!]
\begin{center}
\includegraphics[width=9.0cm,height=10.0cm]{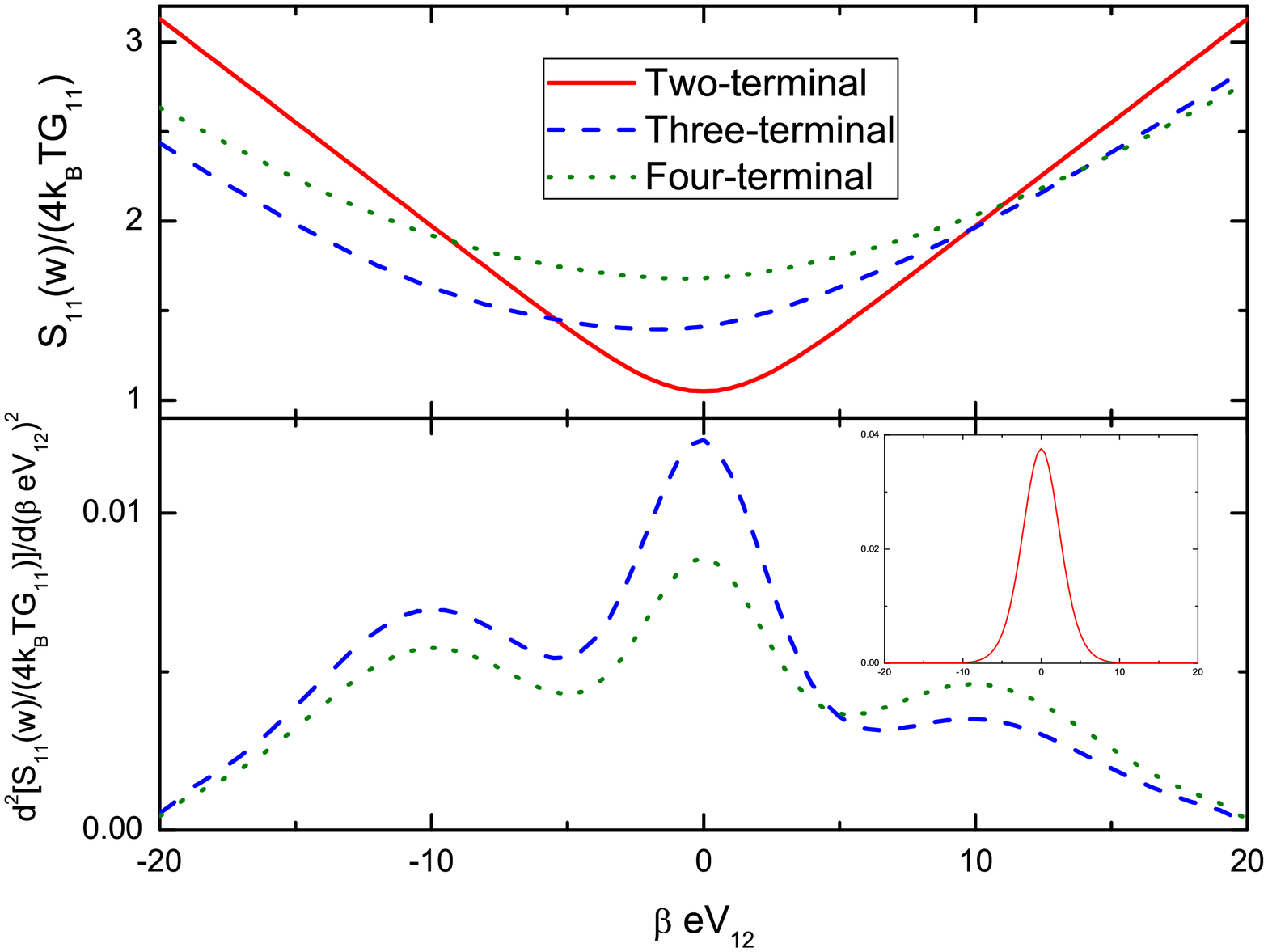}
\end{center}
\caption{The upper panel shows the noise in the regime of finite frequency with capacitive interaction, in terms of $\beta eV_{12}$. The lower panel shows the corresponding second derivative, and the inset depicts its behavior in the case of two terminals. Here we have set $\beta eV_{34}=10$, $\beta \hbar w = 0.3$ and $ \tau/ \hbar\beta = 1$.} \label{frequenciazero}
\end{figure}

\begin{figure}[ht!]
\begin{center}
\includegraphics[width=9.0cm,height=10.0cm]{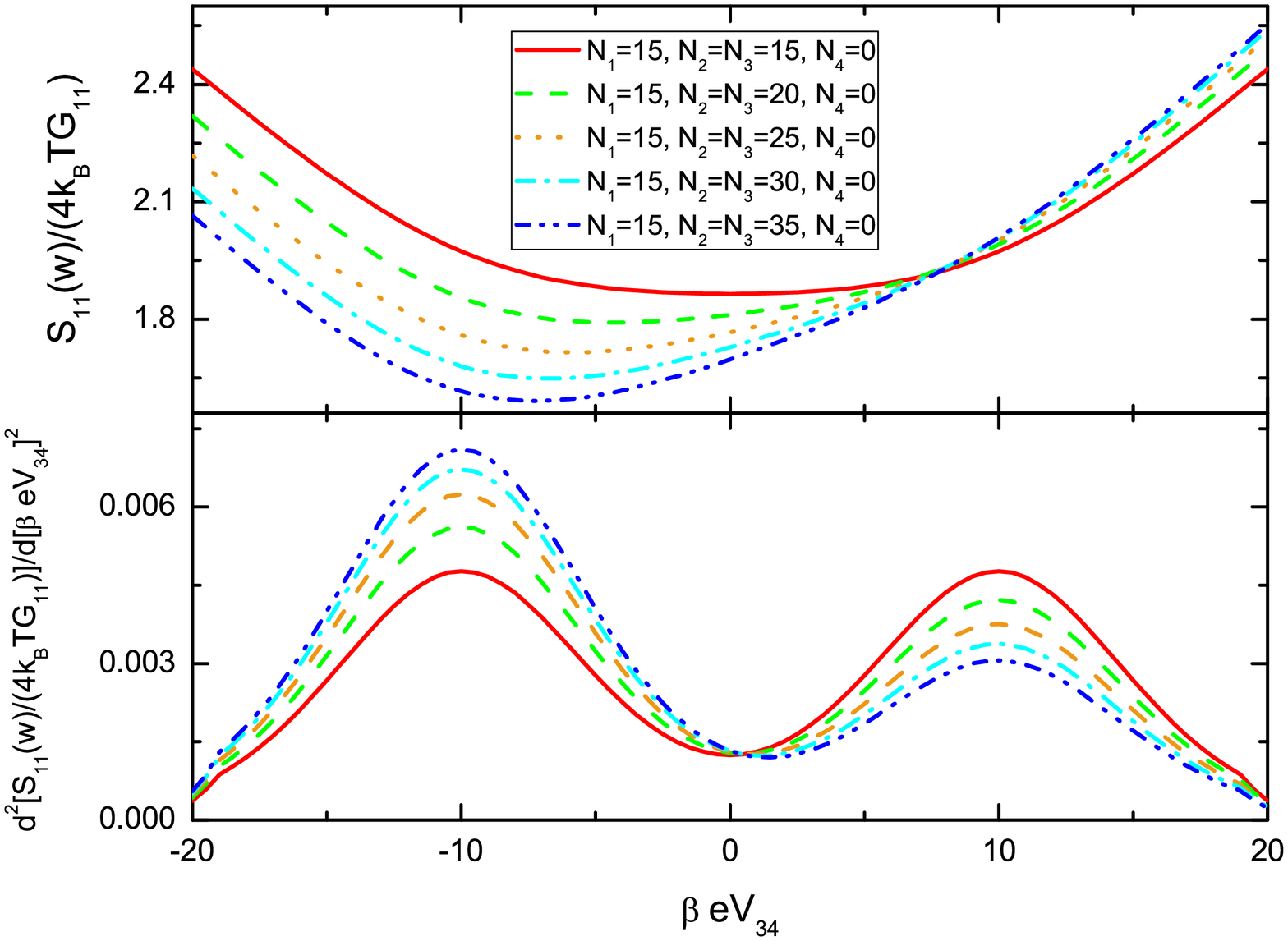}
\end{center}
\caption{The upper panel shows the noise in the regime of finite frequency with capacitive interaction, in terms of $\beta eV_{34}$. The lower panel shows the corresponding second derivative. Here we have set $\beta eV_{12}=10$, $\beta \hbar w = 0.3$ and $ \tau/ \hbar\beta = 1$.} \label{frequenciazero1}
\end{figure}

\section{The Case of Vanishing Frequency}
\label{sistemaquantico1}

Let us now investigate the noise power associated to a chaotic quantum dot connected to four leads, as shown in Fig.~1, in the regime of vanishing frequency.
We first substitute the above results (\ref{main2}) and (\ref{main4}) into Eq.~\eqref{ruido} and then take the limit $w\to0$. Since the general expression is awkward,
we focus the study on some cases of current interest, similar to the ones investigated in \cite{gustavson,pekola}. The first case concerns the regime
where $k_BT,eV_{12}\gg eV_{34}$, in which $V_{34}$, the difference of potential between the leads $3$ and $4$, can be seen as a perturbation.
In this case we can write
\begin{widetext}
\begin{eqnarray}
\mathcal{S}_{ik}(0)&=&\frac{e^2}{h}2k_BT
\frac{N_i(\delta_{ik}M-N_k)}{M^3}(N_1+N_2)(N_3+N_4)\nonumber\\&&
\times\left\lbrace\left[1+\frac{2N_1N_2}{(N_1+N_2)(N_3+N_4)}\right]\frac{eV_{12}}{2k_BT}\coth\left(\frac{eV_{12}}{2k_BT}\right)+\frac{eV_{12}}{2k_BT} \text{csch}\left(\frac{eV_{12}}{2k_BT}\right)\right.\nonumber\\&&\left. +\left[1+\frac{(N_1+N_2)^2+(N_3+N_4)^2-N_1N_2}{(N_1+N_2)(N_3+N_4)}\right]\right\rbrace ,\label{noisefreq011}
\end{eqnarray}
\end{widetext}
where $ i,k=\{1,2,3,4\}$ and $M=\sum_{i=1,}^{4}N_i$. The above result (\ref{noisefreq011}) is in accordance with charge conservation: $\mathcal{S}_{kk}(0)=-\sum_{i=1,i\neq k}^{4}\mathcal{S}_{ki}(0)=-\sum_{i=1,i\neq k}^{4}\mathcal{S}_{ik}(0)$. We can also verify that, if one sets $N_3=N_4=0$, that is, if one closes the channels in the leads $3$ and $4$, we get back to the result recently obtained in Ref.~\cite{epl} for a chaotic quantum dot connected to two leads. Note that the result in the limit $k_BT,eV_{34}\gg eV_{12}$ can be obtained with the exchange of indices $1 \leftrightarrow 3$ and $2\leftrightarrow 4$ in the above result.

Another case of interest can be obtained in the limit $k_BT\gg eV_{12},eV_{34}$. In this regime, we get to the thermal or Johnson-Nyquist noise, which appears in electronic circuits at high temperatures. In this case we get back to the usual relation between noise and conductance, $S_{ik}=4k_BTG_{ik}$, where $G_{ik}$ stands for the conductance between the leads $i$ and $k$. Here the expression   \eqref{noisefreq011} simplifies to
\begin{eqnarray}
{{\mathcal{S}}}_{ik}(0)&=&4k_BT\frac{e^2}{h}\frac{N_i(\delta_{ik}M-N_k)}{M}.\label{condfreq02}
\end{eqnarray}

Let us now analyze the limit  $eV_{12} \gg k_BT$, which leads us to the relevant regime, where the shot noise power is related to the time-dependent fluctuations on the electronic current due to the discreteness of the charge in the regime of low temperatures, $T\to0$. Here we get
\begin{eqnarray}
\mathcal{S}_{ik}(0)&=&\frac{e^3V_{12}}{h}\frac{N_i(\delta_{ik}M-N_k)}{M^3}(N_1+N_2)(N_3+N_4)\nonumber\\&\times&\left[1+\frac{2N_1N_2}{(N_1+N_2)(N_3+N_4)}\right].\label{noisefreq01}
\end{eqnarray}

If we take $N_i=N$ in the above result \eqref{noisefreq01}, we get to the simpler result
\begin{equation}
\mathcal{S}_{ik}(0)=\frac{e^3}{h}\left[\delta_{ik}-\frac{1}{4}\right]\frac{3N}{8}V_{12},
\end{equation}
which agrees with the result given by Eq.~(7) of Ref.~\cite{LangenButtiker}, in the same limit.

\begin{figure}[t!]
\begin{center}
\includegraphics[width=9.0cm,height=10.0cm]{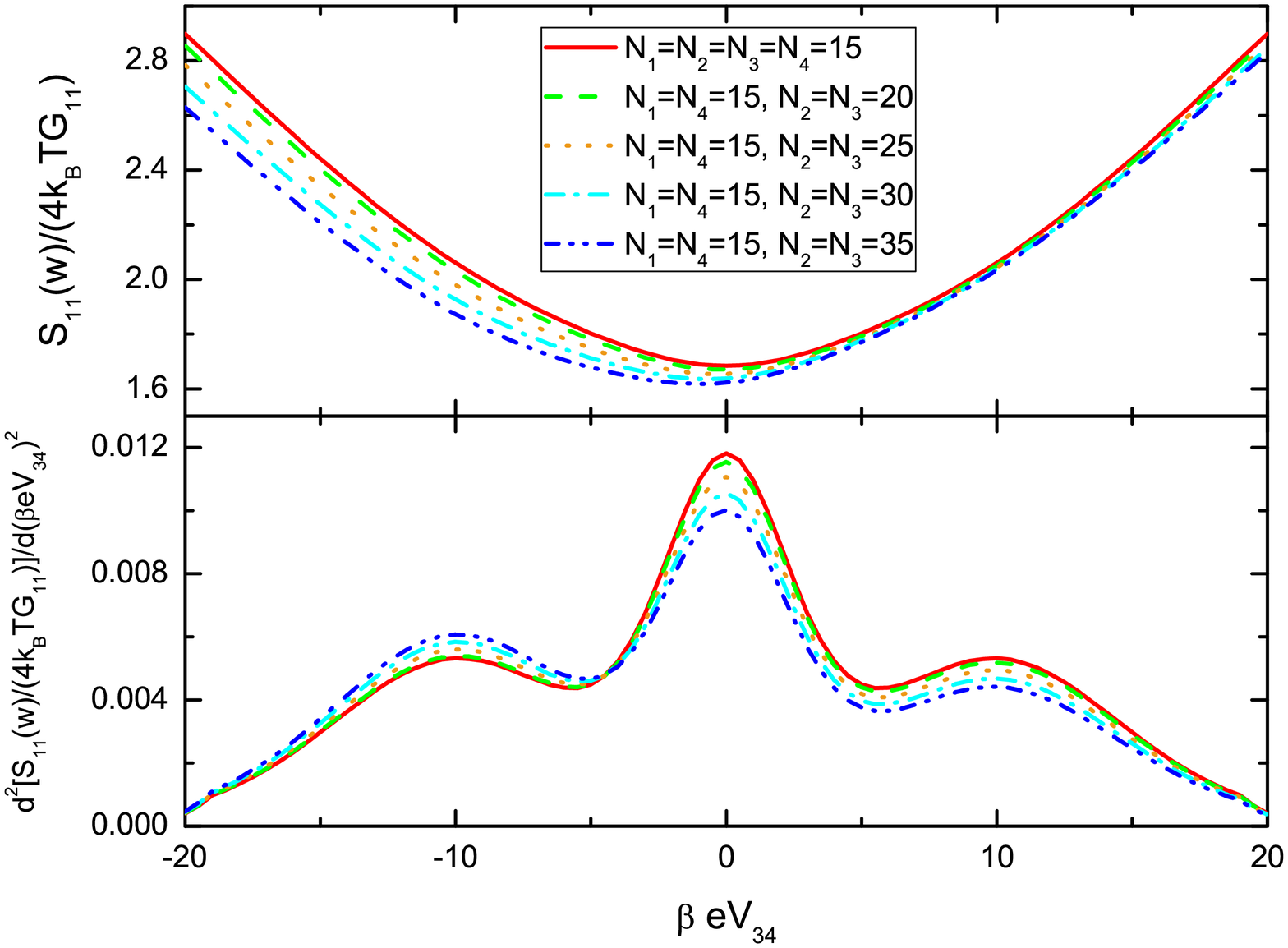}
\end{center}
\caption{The upper pannel shows the noise in the regime of finite frequency with capacitive interaction, in terms of $\beta eV_{34}$. The lower pannel shows the
corresponding second derivative. Here we have set $\beta eV_{12}=10$, $\beta \hbar w = 0.3$ and $ \tau/ \hbar\beta = 1$.} \label{frequenciazero2}
\end{figure}

\begin{figure}[t!]
\begin{center}
\includegraphics[width=9.0cm,height=10.0cm]{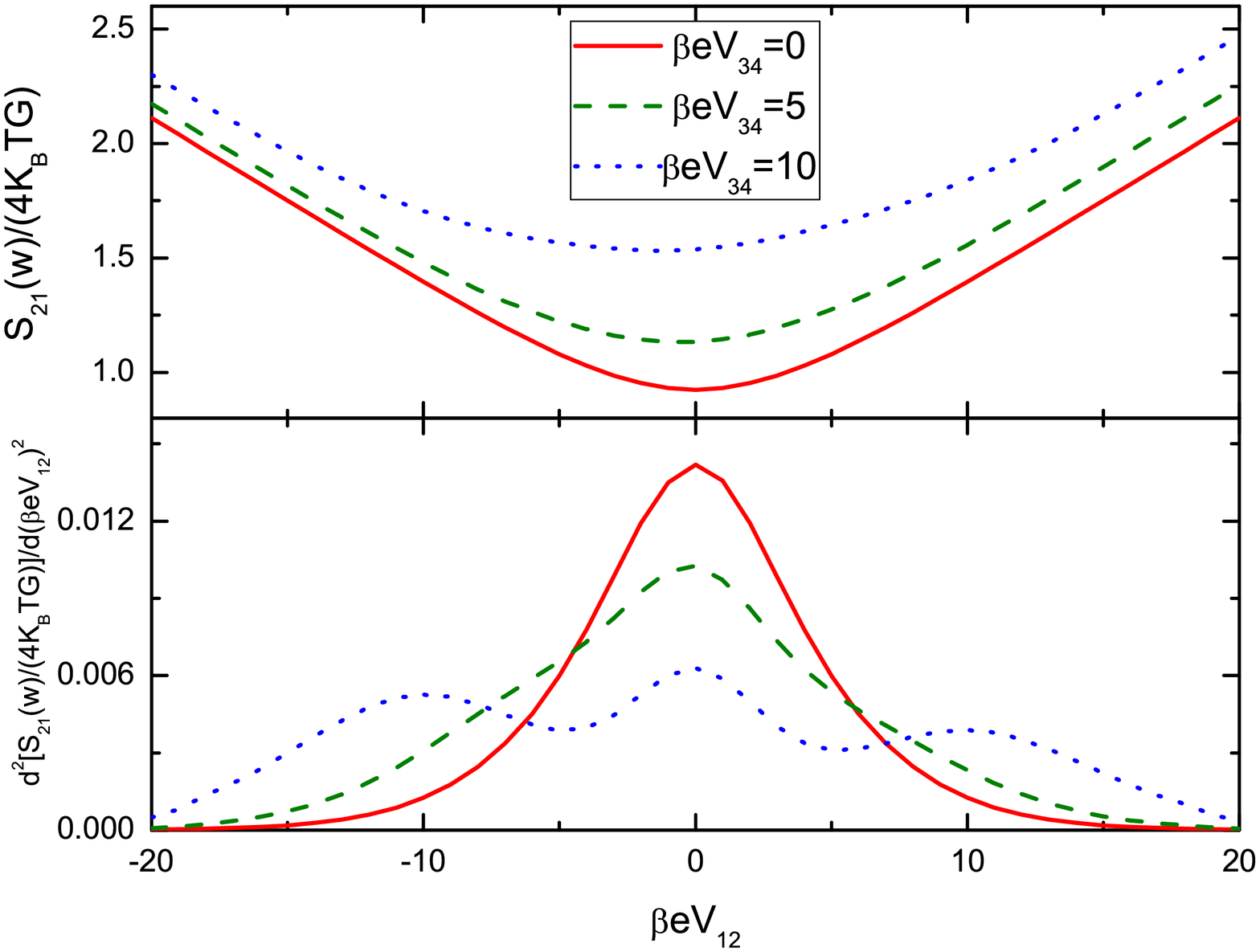}
\end{center}
\caption{Similar plots of the noise $S_{21}$ and its second derivative. Here we have set $N_1=15$, $N_2=40$, $N_3=50$, $N_4=25$, $\beta \hbar w = 0.3$ and $ \tau/ \hbar\beta = 1$.} \label{frequenciazero3}
\end{figure}

\section{Finite frequency and capacitive interaction}
\label{sistemaquantico2}

Let us now study the noise in the regime of non vanishing frequency. As we are going to show, we will need to introduce capacitive interaction in order to ensure charge conservation through the chaotic quantum dot. In this sense, let us first consider the simpler case, where we take the case of nonzero frequency.

\subsection{The Case of Nonzero Frequency}

In order to get to the noise at finite frequency, we have to substitute \eqref{main2} and \eqref{main4} into Eq.~\eqref{ruido}. As before, the final result leads us to an awkward expression. For this reason, we focus on some relevant regimes: firstly, we consider the case where $k_BT\gg eV_{12},eV_{34}$. We get
\begin{eqnarray}
\mathcal{S}_{ik}(w)&=&\frac{e^2}{h}
\frac{N_i(\delta_{ik}M-N_k)}{M}\coth\left(\frac{\hbar w}{2k_BT}\right)\nonumber\\&\times&\frac{2\hbar w\left[1+\delta_{ik}w^2\tau_D^2M/(M-N_k)\right]}{1+w^2\tau_D^2}.\label{ruidofreq1}
\end{eqnarray}
Here we note that $\lim_{w \rightarrow 0}\mathcal{S}_{ik}(w)={{\mathcal{S}}}_{ik}(0)$, where ${{\mathcal{S}}}_{ik}(0)$ is given by the previous result \eqref{condfreq02}, as expected. We also note that after taking $i=k$ in the result \eqref{ruidofreq1}, one verifies the presence of an asymmetry in the exchange among the number of open channels in the leads $N_1, N_2,N_3$ e $N_4$. This asymmetry in the indices of open channels among multiple terminals is similar to the asymmetry found in chaotic quantum dots with two terminals \cite{pekola}.

Another case of interest is given by $eV_{12}\gg k_BT,eV_{34}$. Here we get
\begin{eqnarray}
\mathcal{S}_{ik}(w)&=&\frac{e^3V_{12}}{h}\frac{N_i(\delta_{ik}M-N_k)}{M^3}\nonumber\\
&\times&\frac{1+\delta_{ik}w^2\tau_D^2M/(M-N_k)}{1+w^2\tau_D^2}(N_1+N_2)(N_3+N_4)\nonumber\\
&\times&\left[1+\frac{2N_1N_2}{(N_1+N_2)(N_3+N_4)}\right].\label{ruidofreq3}
\end{eqnarray}
We see that $\lim_{w \rightarrow 0}\mathcal{S}_{ik}(w)=\mathcal{S}_{ik}(0)$, where $\mathcal{S}_{ik}(0)$ is given by the previous result \eqref{noisefreq01}, as expected.
To get to the limit $eV_{34}\gg k_BT,eV_{12}$, in the above result \eqref{ruidofreq1} and \eqref{ruidofreq3}, we change the indices as follows: $1 \leftrightarrow 3$ and $2\leftrightarrow 4$.

Before ending this subsection, let us briefly comment on two interesting issues. The first concerns the presence of asymmetric contribution which appears in the above result \eqref{ruidofreq3}, which will be further considered below. The other issue is that  after taking $N_3, N_4 = 0$ in the above results \eqref{ruidofreq1} and  \eqref{ruidofreq3}, we get back to results obtained in Ref.~\cite{pekola} in the case of two terminals.

\subsection{Presence of Capacitive Interaction}

The noise at finite frequency, as given in \eqref{ruido}, has the symmetry $\mathcal{S}_{ki}(w)=\mathcal{S}_{ik}(-w)$, but it does not ensure current conservation \cite{blanter}.
To remedy the issue and ensure current conservation, according to \cite{PedersenButtiker} we have to take into account the displacement current, induced by the contacts and gates included in the chaotic quantum dot, as shown in Fig.~\ref{ponto}.  We follow this line, and so we substitute the matrix $A_{\beta\alpha}(\varepsilon, \varepsilon')$ by the effective matrix $A_{\beta\alpha}(\varepsilon, \varepsilon')+ \Delta A_{\beta\alpha}(\varepsilon, \varepsilon')$, which is defined as the current matrix plus the displacement current matrix, into
the equation \eqref{ruido}. The displacement current introduces the contribution \cite{PedersenButtiker}
\begin{eqnarray}
\Delta A_{\beta\alpha}(\varepsilon, \varepsilon')&=&2 \pi i\hbar w G_\alpha (w)\nonumber\\&\times &\left[\frac{i}{2\pi}\frac{1_\beta S^{\dagger}(\varepsilon)1_\alpha \left( S(\varepsilon)-S(\varepsilon')\right)}{\varepsilon-\varepsilon'}\right],\nonumber
\end{eqnarray}
where
\begin{eqnarray}
 G_\alpha(w)=\frac{N_\alpha}{M(1-iw \tau)} \frac{\tau_D}{\tau_C+ \tau_D},\nonumber
\end{eqnarray}
is the frequency dependent response function, which was previously obtained in \cite{pekola}, in the case of two terminals, with $\tau$ being the charge relaxation time induced by the displacement current. The charge relaxation time is defined in terms of the dwell time, $\tau_D$, and the RC time, $\tau_C=hC/(Me^2)$, where $C$ is the geometric capacitance of the chaotic quantum dot. It obeys $\tau^{-1}=\tau_D^{-1}+\tau_C^{-1}$.

To obtain the noise in terms of the charge relaxation time, we now have to calculate the average of the trace of
\be
\left(A_{\alpha\beta}(\varepsilon, \varepsilon')+ \Delta A_{\alpha\beta}(\varepsilon, \varepsilon')\right)\left(A_{\beta\gamma}(\varepsilon, \varepsilon')+
\Delta A_{\beta\gamma}(\varepsilon, \varepsilon')\right).
\ee
We substitute the corresponding result into \eqref{ruido} to get to exact expression for the noise power in the case of finite frequency with capacitive interaction. In the limit
of high temperature, $k_BT\gg eV_{12},eV_{34}$, we obtain the following result
\begin{widetext}
\begin{eqnarray}
\mathcal{S}_{ik}(w)&=&\frac{e^2}{h}
\frac{N_i(\delta_{ik}M-N_k)}{M}\coth\left(\frac{\hbar w}{2k_BT}\right)\frac{2\hbar w\left[1+\delta_{ik}w^2\tau^2M/(M-N_k)\right]}{1+w^2\tau^2}\label{ruidofreq22}
\end{eqnarray}
\end{widetext}
Also, in the limit $eV_{12}\gg k_BT,eV_{34}$, we get
\begin{widetext}
\begin{eqnarray}
\mathcal{S}_{ik}(w)&=&\frac{e^3V_{12}}{h}\frac{N_i(\delta_{ik}M-N_k)}{M^3}\frac{1+\delta_{ik}w^2\tau^2M/(M-N_i)}{1+w^2\tau^2}(N_1+N_2)(N_3+N_4)\left[1+\frac{2N_1N_2}{(N_1+N_2)(N_3+N_4)}\right]\label{ruidofreq44}
\end{eqnarray}
\end{widetext}
We note that the above results \eqref{ruidofreq22} and \eqref{ruidofreq44} depend on the charge relaxation time, $\tau$. Thus, we can now ensure that the noise satisfies current conservation through the chaotic quantum dot. It is interesting to note that if we remove the capacitive environment, that is, if we set $\tau_C=0$, we get back to the previous results
\eqref{ruidofreq1} and \eqref{ruidofreq3}. We also note that the limit $eV_{34}\gg k_BT,eV_{12}$ can be obtained from the result \eqref{ruidofreq44}, with the index changes:
$1 \leftrightarrow 3$ and $2\leftrightarrow 4$.

In the next Section, we use the above results to analyze the effects of the number of leads or terminals and the influence of the frequency in the behavior of the noise.

\section{Phenomenology of the noise and asymmetries induced by the extra terminals}

The evaluation of the main term of the noise in a chaotic quantum dot, done with the above generality leads us to large analytical expression of low intuitive capability. To circumvent problem with the understanding of the large analytical expressions, in the current Section we will investigate generic possibilities by means of graphical descriptions, which we believe expose the behavior of the chaotic quantum dot much more significantly. As we shall show, the chaotic quantum dot presents a very rich phenomenology, generated by the competition among the number of terminals, the tension in the reservoirs and the field of finite frequency.

We start with the investigation of the noise as a function of the difference of potential between terminals $1$ and $2$, $\beta e V_{12}$. This is depicted in Fig.~\ref{frequenciazero}, where we used the typical values $N_1=15$, $N_2=40$, $\beta eV_{34}=10$, $\beta \hbar w = 0.3$ e $ \tau/ \hbar\beta = 1$. When the number of open channels in terminals $3$  and $4$ are set to zero,  $N_3=N_4=0$, we see that the noise power is symmetric, with respect to $\beta eV_{12}$. This result corresponds to the system with two terminals, and it is shown in the inset in Fig.~\ref{frequenciazero}. However, we observe an asymmetry when we open channels in the leads $3$ and $4$. We also note that this asymmetry depends on the number of terminals, but it is still present when lead $4$ is closed. To highlight the effect, we also plot the second derivative of the noise power. The plots clearly show the presence of two secondary peaks, also asymmetric.

\begin{figure}[t!]
\begin{center}
\includegraphics[width=9.0cm,height=10.0cm]{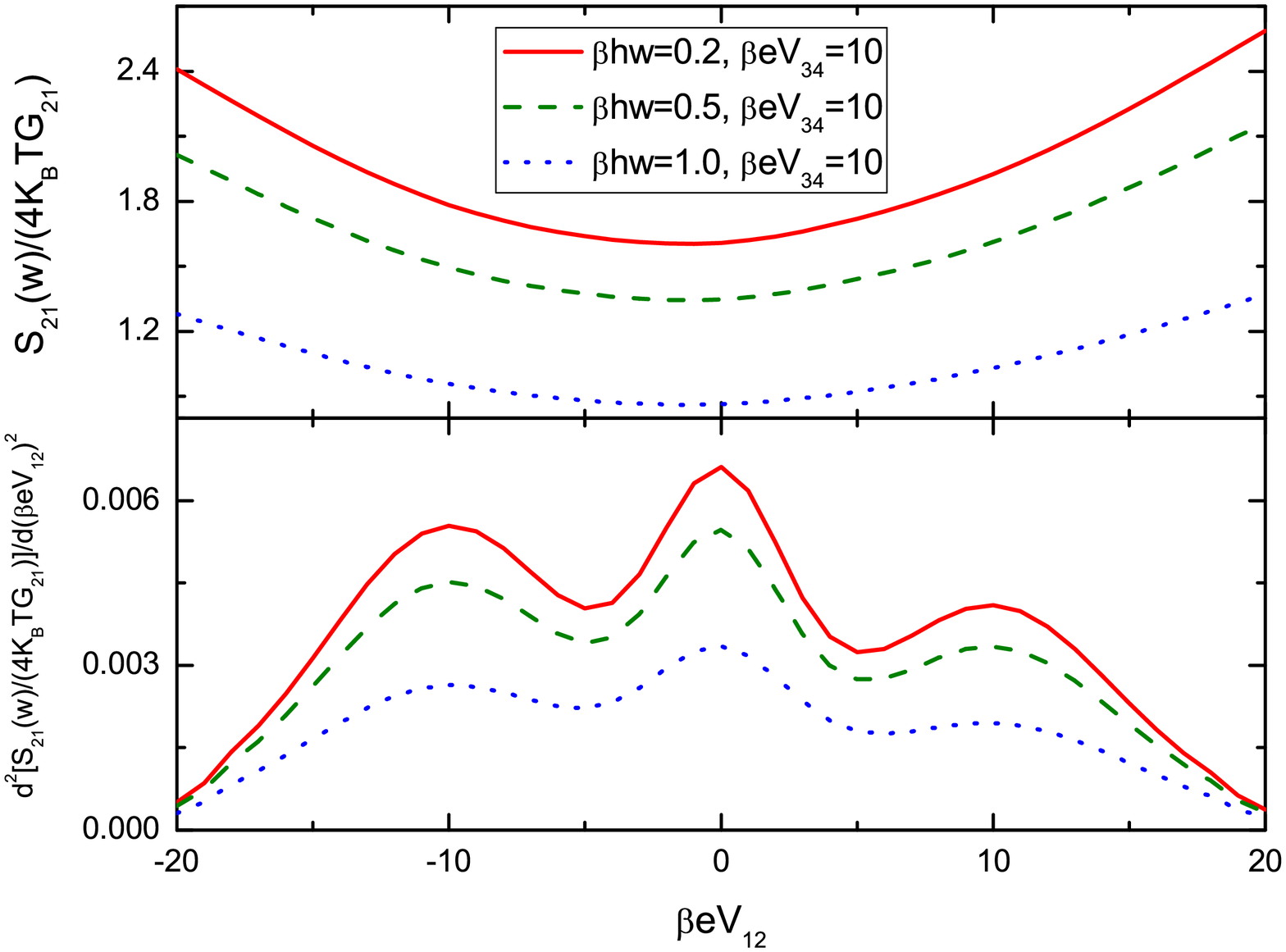}
\end{center}
\caption{Plots of the noise $S_{21}$ and its second derivative. We note a suppression in the peaks of the second derivative of the noise due to the increasing of the frequency. Here we have set $N_1=15$, $N_2=40$, $N_3=50$, $N_4=25$, $\beta eV_{34}=10$ and $ \tau/ \hbar\beta = 1$.} \label{frequenciazero4}
\end{figure}

In Fig.~\ref{frequenciazero1}, we show the symmetry breaking of the noise power, in terms of the difference of potential in terminals $3$ and $4$. There we also show that the symmetry breaking occurs as we vary the number of open channels in the terminals. When the number of open channels in the terminals are symmetric, that is, when one sets $N_1=N_2=N_3$, there is no asymmetry, even though we keep terminal $4$ closed. In a similar way, we show in Fig.~\ref{frequenciazero2} that an asymmetry is also present when there is asymmetry in the number of open channels, in the case with all four leads operating.

In Fig.~\ref{frequenciazero3} we depict competition between $V_{12}$ and $V_{34}$. We note that the second derivative shows the presence of secondary peaks, which increase  for increasing $V_{34}$ and are centered exactly at values where $V_{12}$ equals $V_{34}$, or $V_{43}$. This fact reflects the change of the behavior of the noise due the competition between the potentials. We also note that the noise is symmetric when $V_{12}$ or $V_{34}$ vanishes, and that the asymmetry appears only when they are nonzero, and different from each other.

We have also analyzed the behavior of the noise for different values of $\beta \hbar w$. We illustrate this case in Fig.~\ref{frequenciazero4}, where we show that
the increasing of $\beta \hbar w $ tends to suppress the noise, but the asymmetry due to the potential is still effective.

\begin{figure}[t!]
\begin{center}
\includegraphics[width=9.0cm,height=6.8cm]{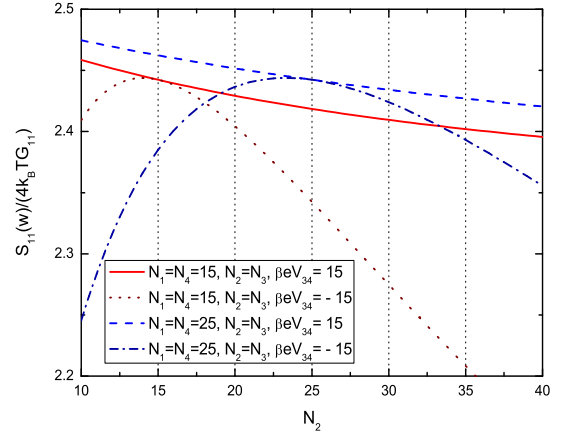}
\end{center}
\caption{Noise in the regime of finite frequency with capacitive interaction, in terms of the number of open channels in terminals $2$ and $3$. Here we have set $\beta eV_{12}=10$,  $\beta eV_{34}=15$, $\beta \hbar w = 0.3$ and $ \tau/ \hbar\beta = 1$.} \label{frequenciazero5}
\end{figure}

Finally, in Fig.~\ref{frequenciazero5} we depict the noise in terms of the number of open channels in the leads. We set $N_1=N_4$, $N_2=N_3$, $\beta eV_{12}=10$, $\beta \hbar w = 0.3$ and $ \tau/ \hbar\beta = 1$, in order to obtain for $S_{11}(\beta eV_{34}=15)$ the approximate value
\ben\label{eq11u}
\frac{10(259N_1^3+1265N_1N_2^2+1023N_1^2N_2+478N_2^3)}{1030N_1^3+4130N_1N_2^2+5160N_1^2N_2+2063N_2^3},
\een
and for $S_{11}(\beta eV_{34}=-15)$, the approximate value
\be\label{eq11v}
\frac{550N_1^3+7110N_1N_2^2+4568N_1^2N_2+1017N_2^3}{452N_1^3+1808N_1N_2^2+2259N_1^2N_2+304N_2^3}.
\ee
In Fig.~ \ref{frequenciazero5} we depict expressions \eqref{eq11u} and \eqref{eq11v} in terms of $N_2=N_3$. They behave differently, and the difference induces the asymmetry of the noise in terms of the number of open channels in the leads. We also note that the two expressions are equal for $N_1=N_2$, meaning that $N_1=N_2$ removes the asymmetry.

\section{Comments and Conclusions}

In this work we studied a chaotic quantum dot connected to four leads in several distinct limits, focusing on the noise at finite frequency and temperature, in the noninteracting and interacting regimes. We used the diagrammatic procedure to obtain analytic results for the main term of the semiclassical expansion for the noise by means of the average over the Wigner-Dyson ensembles. With the diagrammatic procedure, we could write general expressions for the noise. In particular, we studied the noise at lower (shot noise power) and higher (Nyquist-Jonhson noise) temperatures, obtaining general expressions at vanishing or nonvanishing frequency. We also investigated the interacting and non interaction cases, using the capacitive environment under the presence of an AC frequency in the terminals. The main results show the existence of surprising effects induced by the subtle combination of spatial and temporal phase coherence in the electronic modes in the chaotic quantum dot.
Another important issue concerned the topology of the multi-terminal chaotic cavity, which is crucial to modify the behavior of the noise, such as the suppression of the peaks
of its second derivative, and the presence of asymmetry due to nonvanishing tensions applied in the quantum dot. The results also show the importance of the number of open channels in the terminals, to control the behavior of the noise in a multi-terminal chaotic cavity with arbitrary topology.

The study is of current interest and may contribute to understand future experiments on mesoscopic systems with four leads. The effect of topology does not necessarily leads to asymmetry in the conductance, as one can see from the study of Ref.~\cite{petitjean}. However, the quantum nature of the noise in mesoscopic structures appears in a direct manner in the topology. In the very recent work \cite{prlexperimental}, the authors experimentally identify an asymmetry in the noise, induced by the temperature decreasing below the Kondo temperature. The results of the present work are also valid at higher temperatures, so that the asymmetry here identified could be observed in the crossover between the thermal or Nyquist-Johnson noise and the shot noise.

To make the current results stronger, we also investigated the chaotic cavity with six, eight and ten terminals, and the presence of two, four and six extra terminals does not destroy the asymmetry observed in the case of four terminals. We believe that the symmetry present in the case of two terminals is specific to such simple topology, and we hope that our results contribute to the experimental control of the noise in multi-terminal chaotic cavities. There are much more to be done, and we are presently investigating the case of a non ideal chaotic cavity, including the crossover between the orthogonal and unitary ensembles in the weak localization correction, and the role played by the spin accumulation effect in ferromagnetic structures.

\begin{acknowledgments}
This work was partially supported by the Brazilian agencies CAPES, CNPq and FACEPE-PE.
\end{acknowledgments}

\end{document}